\documentclass[letterpaper, 11 pt, conference]{ieeeconf}
\IEEEoverridecommandlockouts 
\usepackage[utf8]{inputenc}
\usepackage{graphicx}                   
\usepackage{listings}                   
\usepackage[frozencache=true, cachedir=.]{minted}                    
\usepackage{color}                      
\usepackage{subfigure}                  
\usepackage{authblk}                    
\usepackage{multirow}                   
\usepackage{multicol}                   
\usepackage{hyperref}                   
\usepackage{caption}                    

\newcommand{\speedup}{2.13x}
\newcommand{\speedupkokkos}{1.36x}

\newcommand\blfootnote[1]{%
  \begingroup
  \renewcommand\thefootnote{}\footnote{#1}%
  \addtocounter{footnote}{-1}%
  \endgroup
}

\title{\LARGE \bf
A Hybrid Graph Coloring Algorithm for GPUs
}

\author[1]{Shanthanu S Rai*}
\author[1]{Rohit M P*}
\author[2]{Sreepathi Pai}

\affil[1]{Department of Computer Science and Engineering, National Institute of Technology Karnataka}
\affil[2]{Department of Computer Science, University of Rochester}
 
\date{Sep 2019}

\begin{document}

\maketitle
\thispagestyle{empty}
\pagestyle{empty}

\begin{abstract}



Graph algorithms mainly belong to two categories, topology-driven and data-driven.
Data-driven approach maintains a worklist of active nodes, the nodes on which work has to be done.
Topology-driven approach sweeps over the entire graph to find active nodes.

Hybridization is an optimization technique where in each iteration, the computation is done in a topology-driven or data-driven manner based on worklist size.
In hybrid implementations, there is a need to switch between topology-driven and data-driven approaches.
Typically, a worklist is maintained just in the data-driven part of the algorithm and discarded in the topology-driven part.
We propose a variant of hybridization, wherein a worklist is maintained throughout all iterations of the algorithm and still show it to be faster than both, topology-driven and data-driven approaches.

We consider a graph coloring algorithm called IPGC (Iterative Parallel Graph Coloring) and implement a hybrid version for the same in a graph domain specific language called IrGL. We observe a mean speedup of \speedup~over a data-driven implementation of IPGC on a suite of 10 large graphs on a NVIDIA GPU.

\end{abstract}

\section{Introduction}
\blfootnote{Note: Student authors have equally contributed towards the paper, order decided by coin toss}

Many applications have problems which can be directly or indirectly mapped to a graph problem.
For example, graphs are used to represent networks which may be a city structure, or a social network, or a telephone network. As such, graph algorithms such as minimum spanning tree and shortest path algorithm find direct uses in the real world.
Intrinsically, graph algorithms suffer from a few problems of their own such as irregular memory access, dynamic data structures, and synchronization bottlenecks, which makes it hard to parallelize these algorithms in a way that utilizes the GPU maximally.

Graph algorithms can be thought of as multiple iterations of applying an operator on a set of nodes~\cite{Pingali:2011:TPA:1993498.1993501}. 
These set of nodes are called active nodes.
This broadly divides graph algorithms into two categories, topology-driven and data-driven.
In topology-driven algorithms, the operator is applied to every node in the graph, and not just the active nodes.
It potentially does unnecessary work because it applies the operator to nodes even when there may be no need for work to be done.
But since the operator is applied to every node, the implementation of topology-driven algorithms is easier because nodes are statically mapped to threads in GPUs. 
Moreover, if most of the nodes are active, then the percentage of wasted work is small, and hence the ease of implementation can be attractive.
Data-driven algorithms, on the contrary, apply the operator only on the list of active nodes.
Thus, they are more work-efficient than topology-driven algorithm.
But they come with the cost of maintaining a list of active nodes per iteration, called a worklist, which uses slow atomic instructions in parallel implementations.

Past work~\cite{Pai:2016:CTO:2983990.2984015,nasre:topo-vs-data} have compared the performance trade-offs between both these approaches and have come up with optimizations specific to each approach.
Nasre et al.~\cite{nasre:topo-vs-data} consider a hybrid optimization that combine both topology-driven and data-driven implementations.
Essentially, based on the size of the worklist, for a given iteration they follow either topology-driven or data-driven approach.
If the worklist size is high, they perform a topology-driven computation and when the worklist size is low, they perform a data-driven optimization.
Thus, they obtain the best of both approaches and eliminate the worst of both.
For switching between data-driven to topology-driven, the worklist of data-driven is simply discarded.
For switching from topology-driven to data-driven, the worklist must be rebuilt.

In this paper, we consider a different type of hybridization wherein the worklist is never discarded.
Even while doing the topology-driven approach, we push active nodes into a worklist.
In section \ref{section:motivation}, we show using a micro-benchmark, the usefulness of this approach and then show an implementation of a graph coloring algorithm wherein the hybrid implementation is \speedup~faster than just a data-driven approach.

\section{Background}
\subsection{Graph Coloring}

Graph coloring is the problem of assigning a color to each node of the graph such there no two endpoints of an edge have the same color, i.e., have no conflicts. 
Of course, we can always assign all nodes with distinct colors. 
Therefore, an additional constraint is to minimize the number of colors used.
While implementing any algorithm to solve this problem, we abstract colors to numbers and assign a number to each node satisfying the ``no conflicts" property.
The problem is proven to be NP-Complete and difficult to approximate~\cite{Zuckerman:2006:LDE:1132516.1132612}. 
Hence, various greedy heuristics have been studied in the literature.

\subsection{IrGL}

IrGL~\cite{Pai:2016:CTO:2983990.2984015} is a compiler that generates CUDA code from an intermediate level program representation.
It also applies throughput optimizations that have been specifically identified to benefit graph algorithms.
In this paper, we use IrGL to implement a parallel graph coloring algorithm called IPGC~\cite{deveci_parallel_2016}.

\subsection{IPGC}

The IPGC algorithm iterates over the following two steps:
\begin{enumerate}
    \item For each uncolored node, assign a color unused by its neighbours. This step could lead to two uncolored nodes sharing an edge to be assigned the same color since the assignments of colors is done in parallel.
    \item If any edge's endpoints have the same color after the previous step, uncolor the affected nodes and go back to step 1.
\end{enumerate}

Assignment of color to a node is done by taking the $mex$\footnote{The $mex$ of a set of positive integers is the smallest value not in the set. For eg., $mex(\{0, 1\}) = 2$, $mex(\{1, 3, 4\}) = 0$} of colors of its neighbours to minimize the number of colors used.
In the data-driven version, we maintain a worklist of conflicting nodes. 
Initially, we assign color 0 to all nodes, implying that they're all uncolored and conflicting. Thus initially, all nodes are put into the worklist.
After each iteration, the number of nodes in the worklist keeps reducing since in case of a conflict, exactly one node from the conflicting edge is removed from the worklist and in case of no conflicts, both the endpoints of the edge are removed thus reducing the size of the worklist by two.
The algorithm terminates when the worklist is empty, i.e., there are no conflicts.

The algorithm is simple, yet uses significantly fewer colors than the NVIDIA-provided CUSPARSE~\cite{Naumov2015ParallelGC} implementation of graph coloring.

\section{Motivation} \label{section:motivation}


Micro-benchmarking is a process that is designed to measure the performance of a specific and small piece of code.
Micro-benchmarks on their own, are not useful in implementing any algorithm, rather they test some specific property of interest and provide performance insights. 
The results obtained from performance of a micro-benchmark inspire optimization techniques that can be used while implementing algorithms.
We use micro-benchmarking to better understand the cost of atomics and the overhead of maintaining a worklist in a data-driven and topology-driven approach.

\begin{listing}
\begin{lstlisting}[
        frame=lines,
        basicstyle=\footnotesize,
        numbers=left,
        framexleftmargin=1.8em,
        xleftmargin=1.8em,
        % numbersep=2pt,
        escapechar=!
    ]
!\textbf{GLOBAL}! COUNT = 1000

!\textbf{Kernel}! Push_WL(graph, iter_count):
    !\textbf{ForAll}! wlnode in WL:
        node = WL.pop(wlnode)
        !\textbf{If}! node> COUNT * iter_count:
            WL.push(node)

!\textbf{Kernel}! Push_NoWL(graph, iter_count):
    !\textbf{ForAll}! node in graph.nodes:
        !\textbf{If}! node > COUNT * iter_count:
            WL.push(node)
\end{lstlisting}
    \caption{Push Kernels}
    \label{lst:pushkernel}
\end{listing}

\begin{listing}
\begin{lstlisting}[
        frame=lines,
        basicstyle=\footnotesize,
        numbers=left,
        framexleftmargin=1.8em,
        xleftmargin=1.8em,
        % numbersep=2pt,
        escapechar=!
    ]
!\textbf{Function}! Exp1:
    iter_count = 0
    WL = graph.nodes
    !\textbf{Pipe}! when size(WL) > 0:
        Push_WL(graph, iter_count)
        iter_count += 1

!\textbf{Function}! Exp2:
    iter_count = 0
    WL = graph.nodes
    !\textbf{Pipe}! when size(WL) > 0:
        Push_NoWL(graph, iter_count)
        iter_count += 1
\end{lstlisting}
    \caption{Kernel Calls}
    \label{lst:exp}
\end{listing}

We consider two such micro-benchmarks, as shown in Listing \ref{lst:pushkernel}. Both the kernels are invoked iteratively within a \texttt{Pipe}~\cite{Pai:2016:CTO:2983990.2984015} as shown in Listing \ref{lst:exp}.
\texttt{Pipe} is an IrGL construct that manages worklists automatically.

We refer to the set of active nodes in the graph, i.e, the set of nodes to be processed in a given iteration, as \texttt{A}.
Kernel \texttt{Push\_WL} iterates over the worklist (data-driven), and \texttt{Push\_NoWL} iterates over the entire graph (topology-driven), both maintaining a worklist all the time and deactivate fixed number of nodes from \texttt{A} in each iteration.
Initially, \texttt{A} contains all nodes.
In each iteration, the first 1000 nodes in \texttt{A} are deactivated. 
So, in the first iteration, nodes labelled 0--999 are deactivated, in the second iteration nodes labelled 1000--1999 are deactivated and so on.

\texttt{Exp1} and \texttt{Exp2} in Listing \ref{lst:exp} invoke the micro-benchmarks.
They initially populate the worklist with all nodes in the graph and then call the micro-benchmarks iteratively within a \texttt{Pipe} until the worklist is empty.

\begin{figure}
    \centering
    \includegraphics[width=\linewidth]{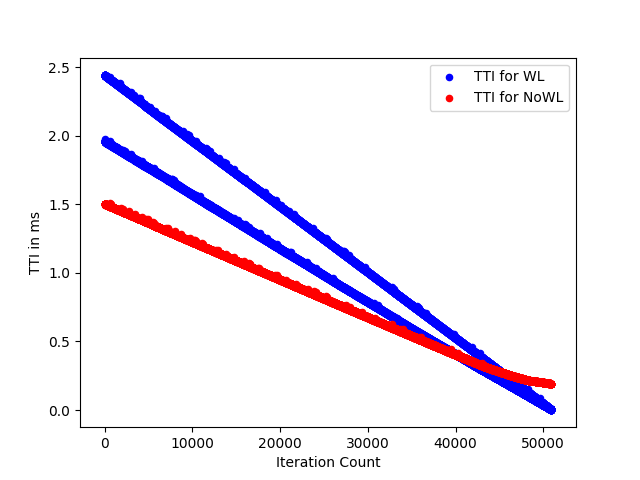} 
    \captionsetup{justification=centering}
    \caption{Time taken per iteration (TTI) of the kernels in Listing~\ref{lst:pushkernel} averaged over 10 runs.}
    \label{fig:PushPlot}
\end{figure}

We run the functions in Listing \ref{lst:exp} on europe\_osm graph on a NVIDIA GPU. Figure \ref{fig:PushPlot} plots the the time taken for each iteration (abbreviated to ``TTI") of the \texttt{Pipe} for both \texttt{Exp1} and \texttt{Exp2} averaged over 10 runs.
Observe that the TTI for \texttt{Push\_NoWL} mostly lie along a line whereas, for \texttt{Push\_WL} TTI for each iteration shows bimodal behaviour and don't necessarily lie along one line.
    
Thus, although both the micro-benchmarks differ only in what they iterate over, there is significant difference in terms of TTI.
We define points where the TTI of \texttt{Push\_NoWL} cross the TTI of \texttt{Push\_WL} as crossover points.
This observation of changes in TTI based on the size of the list of active nodes for simple topology-driven and data-driven implementations hint towards a hybrid optimization technique.

Note that both the micro-benchmarks we defined do the same work, i.e., deactivate the same subset of 1000 nodes from the set of active nodes in any given iteration.
Now suppose we want to optimize this task. 
From Figure~\ref{fig:PushPlot}, we observe that \texttt{Push\_NoWL} is faster than \texttt{Push\_WL} for the initial computation upto the first crossover point, approximately at iteration count of 40000.
After this point, we observe that \texttt{Push\_WL} becomes faster.
Note that as we progress in iterations, the worklist size strictly decreases.
So an ideal hybrid implementation, that picks the faster of \texttt{Push\_NoWL} and \texttt{Push\_WL} would be faster than both \texttt{Exp1} and \texttt{Exp2}.
A real implementation could use the size of the worklist to switch between the two.


\section{Hybridization}


We can speedup data-driven algorithms by switching to topology-driven (but still maintain the worklist) when the number of active nodes is high.
The switch is cheap since it is based just on size of the worklist.

We propose a hybridization technique that can be applied to any graph algorithm.
Following are the steps to implement a hybrid version of a algorithm:

\begin{enumerate}
    \item Write two kernels which implement the data-driven approach and the topology-driven approach for the algorithm while maintaining a worklist all the while. Figure \ref{fig:topo-from-data} shows the differences between the two:    
    in the data-driven version, the outer loop iterates over the worklist and active nodes are read from the worklist whereas in the topology-driven version, the outer loop iterates over all graph nodes and active nodes are determined in an algorithm-specific way.
    
    \begin{figure}
        \centering
        \captionsetup{justification=centering}
        \includegraphics[width=\linewidth]{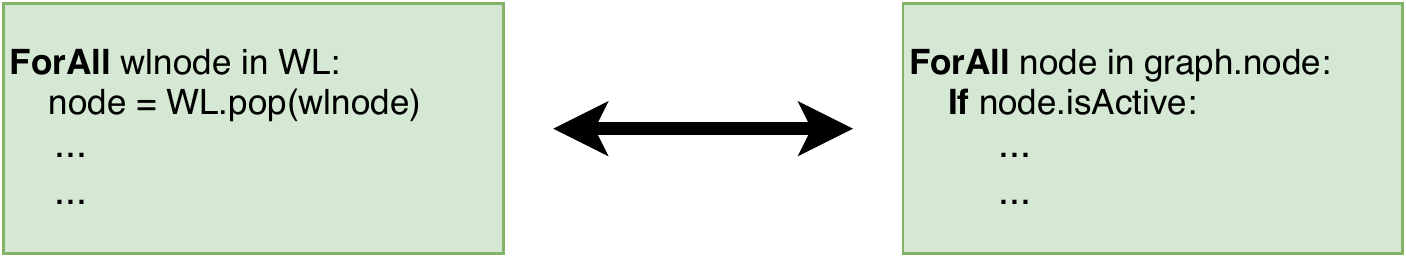}
        \caption{Differences between data-driven and topology-driven kernel}
        \label{fig:topo-from-data}
    \end{figure}
    

    \item Create a driver function that decides which type of kernel to call based on worklist size. Let \texttt{H} be the threshold value of worklist size based on which kernels to invoke is decided. Note that \texttt{H} is a tuning parameter. Figure \ref{fig:driver-function} shows the control diagram for the driver function. 
    
    \begin{figure}
        \centering
        \includegraphics[width=\linewidth]{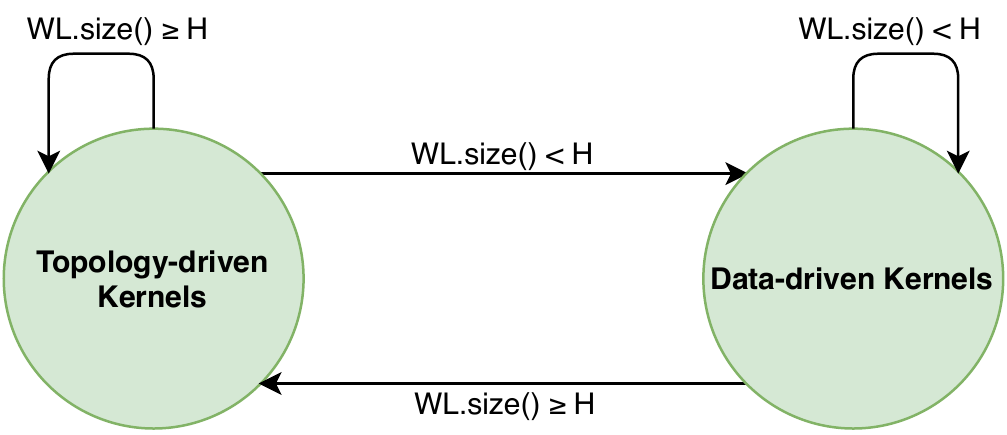}
        \caption{Control algorithm of hybridized driver function}
        \label{fig:driver-function}
    \end{figure}
    
\end{enumerate}


We experimentally determined that a value of \texttt{H} around 60\% of the total number of graph nodes provides the best speedup for the 10 graphs in Table \ref{table:graphs} on the NVIDIA Quadro P5000 for IPGC.

\begin{table}
\fontsize{7}{10}\selectfont
\centering
\begin{tabular}{||l|l|l|l|l|l||}
    \hline
    Graph & Nodes & Edges & $\delta_{min}$ & $\delta_{median}$ & $\delta_{max}$       \\
    \hline \hline
    circuit5M & 5.5M & 53.9M & 0 & 4 & 1290500              \\
    \hline
    Audikw\_1 & 0.9M & 76.7M & 20 & 68 & 344                \\
    \hline
    Bump\_2911 & 2.9M & 124.8M & 0 & 44 & 194               \\
    \hline
    Queen\_4147 & 4.1M & 325.3M & 23 & 80 & 80              \\
    \hline
    kron\_g500-logn21 & 2.0M & 182.1M & 0 & 4 & 213904      \\
    \hline
    indochina-2004 & 47.4M & 302.0M & 0 & 11 & 256425       \\
    \hline
    hollywood-2009 & 1.1M & 112.8M & 0 & 28 &11467          \\
    \hline
    rgg\_n\_2\_24 s0 & 16.8M & 265.1M & 0 & 16 & 40         \\
    \hline
    soc-LiveJournal1 & 4.8M & 85.7M & 0 & 5 & 20333         \\
    \hline
    europe\_osm & 50.9M & 108.1M & 1 & 2 & 13               \\
    \hline
\end{tabular}
\caption{Graph and Degree ($\delta$) Statistics}
\label{table:graphs}
\end{table}

\section{Results}


We compare the performance of various implementations of IPGC against our proposed hybrid implementation. 
We evaluate the performance on a NVIDIA Quadro P5000, with CUDA version 10.0 and CUDA driver version 410.79.
Table \ref{table:versions} shows the four implementations we consider for comparison.
We evaluate each implementation based on two metrics, the time taken and the chromatic number (the number of colors used in the coloring).
We consider 10 input graphs for coloring as shown in table \ref{table:graphs}. 
All the graphs were used by Deveci et al.~\cite{deveci_parallel_2016} and were obtained from UFL Sparse Matrix Collection~\cite{Davis:2011:UFS:2049662.2049663}. 
Graphs were pre-processed to remove self loops and multiple-edges between nodes.

The circuit5M and indochina-2004 graphs are highly irregular with widely varying node degree, while Audikw\_1, Bump\_2911, Queen\_4147, rgg\_n\_2\_24\_s0 and europe\_osm are very regular graphs. 
Specifically europe\_osm has a very low average degree of 2 and is very regular.

\begin{table}
\fontsize{7}{10}\selectfont
\centering
\begin{tabular}{ || l | l || }
    \hline
    Version     & Description                           \\
    \hline \hline
    CUSPARSE    & Implementation using CUSPARSE Library \\
    \hline
    Kokkos      & Implementation using Kokkos Library   \\
    \hline
    Plain       & Implementation in IrGL                \\
    \hline
    Hybrid      & Hybridization in IrGL                 \\
    \hline
\end{tabular}
\caption{Various implementations of IPGC}
\label{table:versions}
\end{table}

Kokkos~\cite{deveci_parallel_2016} and CUSPARSE~\cite{Naumov2015ParallelGC} are state of the art implementations. 
During evaluation of their performance, we encountered certain problems with CUSPARSE and Kokkos.
CUSPARSE yielded an invalid coloring, with 3 conflicts for the circuit5M graph. 
Kokkos provides a converter to convert graphs in .mtx format (as available in UFL Sparse Matrix Collection) to  specific .bin format. 
The converter failed to convert europe\_osm from .mtx to .bin format. We do not show the speedup results for these graphs.

Both the Plain and Hybrid IrGL versions were optimized using Iteration Outlining, Cooperative Conversion and Nested Parallelism~\cite{Pai:2016:CTO:2983990.2984015}.

\begin{figure}
    \centering
    \includegraphics[width=\linewidth]{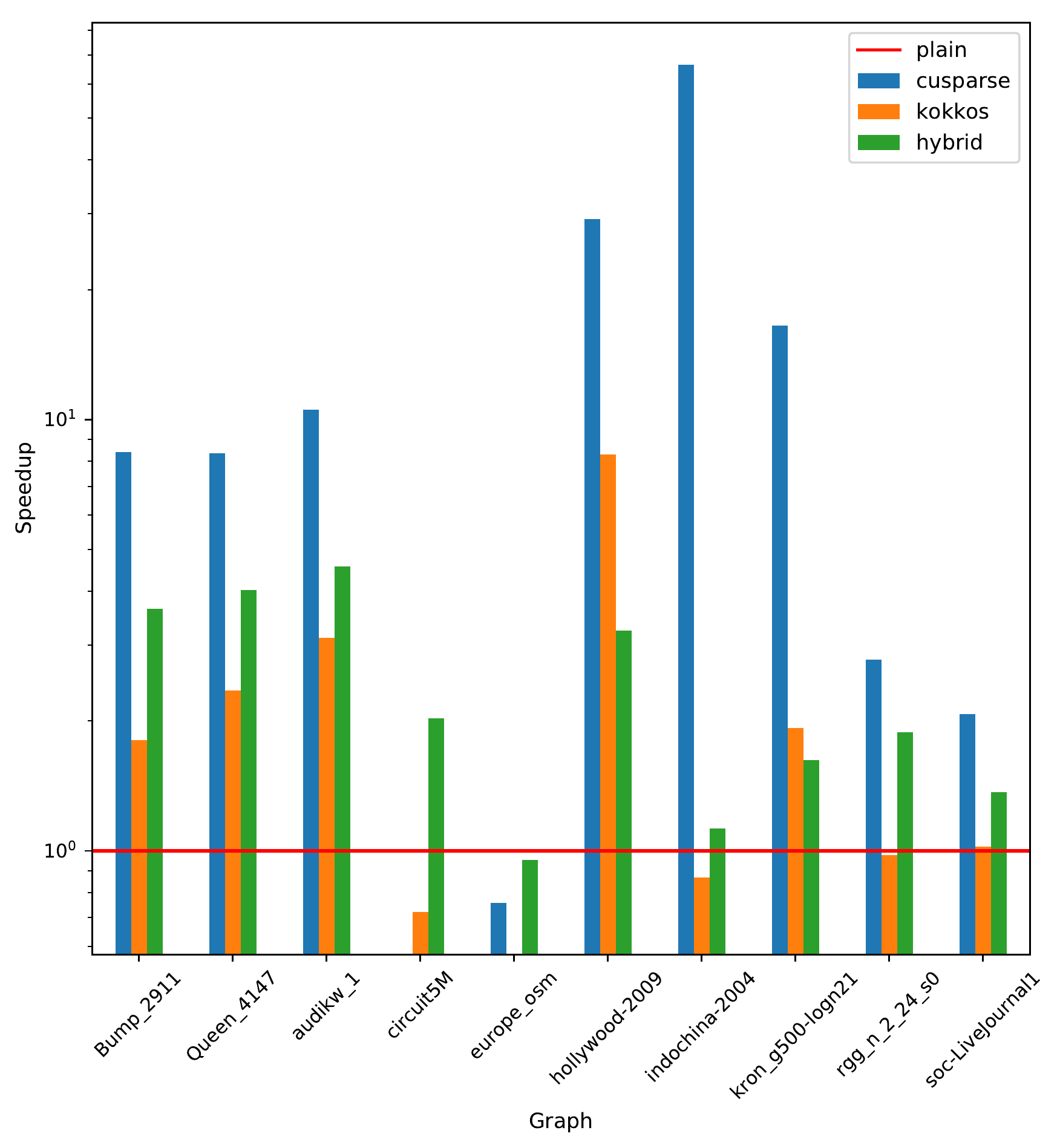}
    \caption{Speedup graph}
    \label{fig:speedup}
\end{figure}

Figure~\ref{fig:speedup} plots of speed up of Kokkos, CUSPARSE and hydridized version over Plain version of IrGL.
Table~\ref{table:times} has the absolute values for the same figure.
We notice that for most graphs CUSPARSE vastly outperforms every other version.
But from Table~\ref{table:colorsUsed} we see that though CUSPARSE is much faster, it uses many more colors than the other versions.
In Table~\ref{table:colorsUsed} we don't mention colors utilised by Plain and Kokkos version because it is the same as Hybrid, since they all implement exactly the same algorithm for assigning colors, just with different optimizations.

\begin{table}
\fontsize{7}{10}\selectfont
\centering
\begin{tabular}{||l|l|l|l|l||}
    \hline
    Graph & CUSPARSE & Kokkos & Plain & Hybrid \\
    \hline \hline
    Bump\_2911 & 69.67 & 324.19 & 586.42 & 160.88 \\
    \hline
    Queen\_4147 & 204.14 & 724.08 & 1705.68 & 423.76 \\
    \hline
    Audikw\_1 & 56.60 & 191.62 & 597.55 & 130.84 \\
    \hline
    circuit5M & - & 2606.41 & 1879.49 & 925.56 \\
    \hline
    europe\_osm & 401.58 & - & 304.17 & 319.56 \\
    \hline
    hollywood-2009 & 526.16 & 1844.44 & 15321.88 & 4723.84 \\
    \hline
    indochina-2004 & 2776.43 & 212393.19 & 184496.37 & 163778.19 \\
    \hline
    kron\_g500-logn21 & 4597.60 & 39475.64 & 76039.89 & 46879.66 \\
    \hline
    rgg\_n\_2\_24\_s0 & 204.12 & 579.76 & 566.50 & 300.85 \\
    \hline
    soc-LiveJournal1 & 194.89 & 394.65 & 404.11 & 295.84 \\
    \hline
\end{tabular}
\captionsetup{justification=centering}
\caption{Time taken (in ms) for all versions (averaged over 3 runs)}
\label{table:times}
\end{table}

Amongst Plain, Kokkos and Hybrid, Hybrid does the best for most graphs. It has an average speedup (geometric mean) of \speedup~over the Plain version and an average speedup of \speedupkokkos~over the Kokkos implementation.

\begin{table}
\fontsize{7}{10}\selectfont
\centering
\begin{tabular}{||l|r|r||}
    \hline
    Graph & Hybrid & CUSPARSE           \\
    \hline \hline
    circuit5M & 8.0 & 341.0*          \\
    \hline
    Audikw\_1 & 56 & 160.0            \\
    \hline
    Bump\_2911 & 32.4 & 96.0           \\
    \hline
    Queen\_4147 & 48.2 & 128.0         \\
    \hline
    kron\_g500-logn21 & 806.0 & 3465.0   \\
    \hline
    indochina-2004 & 6856.8 & 7030.0     \\
    \hline
    hollywood-2009 & 2208.2 & 2317.0      \\
    \hline
    rgg\_n\_2\_24 s0 & 24.0 & 64.0   \\
    \hline
    soc-LiveJournal1 & 340.4 & 557.0     \\
    \hline
    europe\_osm & 5.0 & 32.0        \\
    \hline
\end{tabular}
\caption{Average number of colors used}
\label{table:colorsUsed}
\end{table}

\section{Related Work}

To the best of our knowledge, this is the first attempt at hybridization wherein the worklist is maintained throughout all iterations of the algorithm, i.e., in both topology-driven and data-driven parts. 


Beamer et al. \cite{beamer:dir:bfs} considers a hybrid version of BFS, where they hybridize between a conventional top-down and proposed bottom-up approach. The top-down approach is faster than bottom-up when the BFS frontier size is small and vice versa. For the top-down approach they maintain a queue (worklist) and for bottom-up approach they maintain a bitmap and convert between both data structures when they switch between the two approaches.

Nasre et al. \cite{nasre:topo-vs-data} introduces temporal and spatial hybridization. Temporal hybridization uses data-driven and topology-driven approaches based on the worklist size. In their work, while switching from data-driven to topology-driven, the worklist is discarded, and while switching from topology-driven to data-driven, the worklist is rebuilt.

\section{Future Work}

The results obtained and shown demonstrate the usefulness of our hybridization technique for one specific graph algorithm -- the IPGC graph coloring algorithm.
We will apply this technique to other graph algorithms in future work.

Currently, \texttt{H}, the threshold for switching is determined empirically.
We intend to develop analytical techniques in the future.
Finally, we do not completely understand why topology-driven approach is better than the data-driven approach for some iterations even though both maintain a worklist, and this requires further study.

\section{Conclusion}

We have demonstrated that our hybridization technique wherein we maintain a worklist throughout the computation was \speedup~faster than the data-driven IPGC algorithm and was 36\% faster than Kokkos on average.




\bibliographystyle{plain}
\bibliography{paper}

\end{document}